\documentclass[onecolumn,aps,floatfix,superscriptaddress]{revtex4-2}

\usepackage{color}
\usepackage{graphicx}
\usepackage{bm}
\usepackage{amsmath}
\usepackage{amsbsy}
\usepackage{hyperref}
\usepackage{enumerate}
\usepackage{upgreek}
\usepackage[subnum]{cases}
\usepackage{dsfont}

\begin{document}

\title{Surface diffusion: The intermediate scattering function seen as a characteristic function of probability theory}
		
\author{E. E. Torres-Miyares}
\email{elena.torres@iff.csic.es}
\affiliation{Instituto de F\'isica Fundamental, Consejo Superior de Investigaciones Cient\'ificas, Serrano 123, 28006 Madrid, Spain}

\author{S. Miret-Art\'es}
\email{s.miret@iff.csic.es}
\affiliation{Instituto de F\'isica Fundamental, Consejo Superior de Investigaciones Cient\'ificas, Serrano 123, 28006 Madrid, Spain}

\begin{abstract}
	
In surface diffusion, one of the key observables is the so-called intermediate scattering function which is measured directly from the surface technique called Helium spin echo. In this work, we show that this function can be seen as a characteristic function of probability theory. From the characteristic function, the moments and cumulants of the probability distribution function of the position of the adsorbate are straightforward obtained in an analytical way; in particular, the second order which is related to the diffusion coefficient. In order to illustrate this simple theory, we have focused on the incoherent tunneling of H and D on a Pt(111) surface where only jumps between nearest neighbor sites have been reported experimentally. Finally, an extension to jumps to more than nearest neighbors has also been considered.  
\end{abstract}
	
\maketitle

%

Diffusion of adsorbates/adparticles on surfaces sampled by He atoms instead of neutrons is a very active field due to the fact that diffusion is one of the primary processes occurring on surfaces. Two well established surface experimental techniques are widely used in this context, the quasi-elastic He atom scattering (QHAS), \cite{Hofman1996,Graham2003} and neutron scattering (QENS) \cite{Irene2012} which have been complemented by using the spin-echo (SE) technique, HeSE, \cite{Jardine2009a,Jardine2009b} and neutron spin-echo NSE. \cite{Irene2012}  Unlike neutron scattering, helium scattering is fully coherent. \cite{ward2021inter} In QHAS experiments, the so-called differential reflection
coefficient $\mathcal{R}$ (or the probability that the 
probing particles are scattered when interacting 
with adsorbates), forming a certain solid angle $\Omega$ with 
an energy exchange $\hbar \omega$ and a  momentum transfer parallel to surface, $\textbf{K}$, is proportional to the so-called dynamic structure factor (DSF), $S(\textbf{K},\omega)$, or response function according to
\begin{equation}\label{R}
	\frac{d \mathcal{R}^2 (\textbf{K},\omega)}{d \Omega \, \, d (\hbar \omega)}  = n_d \, F^2 \, S(\textbf{K},\omega),
\end{equation}
where $n_d$ is the diffusing particle concentration on the 
surface, $F$ is the atomic form factor which depends on 
the mutual interaction between probe particles and $N$ interacting adsorbates. Here, capital letters are often used to design variables on the surface. The DSF also gives the spectrum of spontaneous fluctuations. 

On the other hand, in HeSE experiments, the response function is the intermediate scattering function (ISF), $I(\textbf{K},t)$, which is also obtained through the DSF by 
a frequency Fourier transform as follows
\begin{equation}\label{ISF1}            
I(\textbf{K},t)  =  \frac{1}{2 \pi} \int d \omega \, e^{ i \omega t} \, S(\textbf{K},\omega)  .
\end{equation}
In neutron scattering, \cite{Lovesey} the ISF is the Fourier transform of the autocorrelation function of the particle density operator in the reciprocal space with 
$\rho_\textbf{K}(t) = \sum_{j=1}^N \exp(- i \,\textbf{K} \cdot \textbf{R}_j(t))$; $\textbf{R}_j(t)$ being the position operator  of the adsorbates as a function of time. Within the Born approximation, 
$S$ and $I$ can be expressed in terms of the generalized pair-distribution function, also known as the van Hove space-time correlation function. This function is the probability that given 
a particle at the origin at time $t=0$, any particle, including the same one, can be found at the position $\textbf{R}$ and at time $t$.

In order to extract relevant physical information about the system of interest, several theoretical methods have been widely used. Thus, molecular dynamics calculations are generally used where a full description of the force fields (adsorbate-adsorbate and adsorbate-substrate interactions) involved is necessary, Langevin calculations by including or not memory effects, reduced density matrix calculations within the Caldeira-Leggett (CL) and Lindblad formalisms. Except in the first type of calculations, the surface is usually well represented by a thermal bath consisting of an infinite number of harmonic oscillators and, therefore, friction and noise (white or color) appear after integrating over the degrees of freedom of the surface. 
An equation of motion is then derived for the time evolution of the reduced density matrix which contains both frictional and thermal effects, the so-called CL equation \cite{Caldeira1983} which is of Markovian type. The corresponding diagonal matrix elements give the quantum probabilities and the off-diagonal elements, the coherences. The same happens in the Lindblad formalism.\cite{Lindblad}
The Lindblad master equation \cite{lindblad1976generators,Kossakowski} is usually solved numerically through the stochastic wave function method.  \cite{breuer2002theory,torres2021surface} In any case, the ISF can thus be expressed as \cite{torres2021surface,Torres2022,Torres2023,Torres2025} 
\begin{equation}\label{ISF2}            
I(\textbf{K},t)  =  \int d \textbf{R} \, e^{ i \textbf{K}\cdot\textbf{R}} \, \rho(\textbf{R},t) = \langle  e^{ i \textbf{K}\cdot\textbf{R}(t)} \rangle ,
\end{equation}
where $\rho(\textbf{R},t)$ represents the diagonal elements of the reduced density matrix.

Interestingly enough, the second part of Eq. (\ref{ISF2}) is indicating us that the ISF can also be seen as a characteristic function (in this case, depending on time) coming from probability theory.\cite{Gardiner,Haag} A characteristic function is defined as the Fourier transform of the probability distribution function fulfilling the well-known properties: (i) $I(\textbf{K},0)= 1$, (ii) $I(\textbf{K},t)\leq I(\textbf{K},0)= 1$, (iii) $I(\textbf{K},t)$ is a uniformly continuous function of its arguments for all $\textbf{K}$ values, (iv) the moments of the probability distribution are obtained by Taylor expansion of the exponential function and, therefore, one can speak of the momentum generating function, (v) the natural log of the ISF is the corresponding cumulant generating function, and (vi) the Fourier inversion formula exists determining the probability distribution. Furthermore, the properties of convergence, independent random variables and the sum of independent random variables are also fulfilled.

As an application of this new approach, we are going to consider the incoherent tunneling of H/D on a Pt(111) surface analyzed by means of the He spin-echo  (HeSE) scattering experimental technique.\cite{jardine2010determination} Diffusion dynamics in one direction (say, coordinate X) was claimed to take place in the moderate-to-high friction regime.
Quantum effects were reported to be significant at low temperatures, covering a range of surface temperatures from 250 K up to 80 K and therefore thermal activation and tunneling regimes coexist. The crossover temperature was estimated to be 66 K for H and 63 K for D. Diffusion motion seems to correspond to nearest neighbor hopping for a coverage of $0.1$ ML, along the $[11\bar2]$ direction and with a lattice length of $a=2.77 \, \text{\AA}$. Following the well-known Chudley-Elliott (CE) model,\cite{Chudley} deviations from nearest neighbor random jumps for H and D between fcc hollow sites were reported to be minimal.

Within the formalism of the master equation (the CE model can be considered as a special case), \cite{Haag,Weiss} if a simple Bravais lattice is assumed as well as instantaneous jumps between adjacent sites/wells (the diffusion dynamics is assumed to be one-dimensional along a symmetry direction on a periodic substrate, coordinate $X$), a Pauli master equation can be written in terms of probabilities to find the adsorbate on a given potential well $n$, $P_n(t)$, as \cite{Ruth2013,Torres2025}
\begin{equation}\label{Pdot1}
	\dot P_n (t) = \Gamma^+_{n-1} P_{n-1}(t) +  \Gamma^-_{n+1} P_{n+1} (t) - (   \Gamma^+_{n} + \Gamma^-_{n} )  P_{n}(t)  ,
\end{equation}
with $\Gamma^{\pm}_{n \pm 1}$ being the tunneling/hopping transition rates from the $(n \mp 1)$-th well to the $n$-th well and $\pm$ denotes if diffusion goes to the right or to the left, respectively. Usually, these transition rates are fitting parameters or can be calculated from a theory; for example, Kramers' theory. \cite{Eli2023} If the initial condition is such that $P_n(0) = \delta_{n0}$ and $\Gamma = \Gamma^+ + \Gamma^-$ describes the total rate (with $\Gamma^+_{n} = \Gamma^-_{n}$,  $\Gamma^+_{n} = \Gamma^+$, and $\Gamma^-_{n} = \Gamma^-$). By solving Eq. (\ref{Pdot1}), one obtains 
\begin{equation}\label{Psol}
	P_n (t) = I_n(\Gamma t) \, e^{- \Gamma t},
\end{equation}
where $I_n(x)$ is the modified Bessel function of integer order $n$.  Thus, $P_n(t)$ gives us then the probability to stay in the $n$th-well of the binding site at time $t$. Then, the ISF can then be written as
\begin{equation}\label{Ik}
	I (K,t) = \sum_{n= - \infty}^{n= + \infty} P_n(t) \, e^{i K_{||} n} = e^{- \Gamma t} \sum_{n=- \infty}^{+ \infty} I_n(\Gamma t) e^{i K_{||}n } 
    = e^{- \Gamma  (1-cos \, K_{||}) t} ,  
\end{equation}
where $K_{||} = K a \cos(\beta)$, $\beta$ being the angle formed by the direction of observation and diffusion symmetry direction (in our case, $\beta = \pi / 6$ for the two equivalent symmetry directions allowed by $\text{\bf K}$). 

On the other hand, in the so-called diffusive time regime (when time is much greater than the inverse of the friction), it is generally assumed that the ISF  can be fitted by an exponential function of time according to
\begin{equation}\label{eq:I}
I({\bf K}, t) = B \, e^{-   \alpha({\bf K}) \, t} + C,
\end{equation}
where $B$ and $C$ are constants for a given $\textbf{K}$, which is the momentum transfer between the probe particle and the adsorbate along the surface, and $\alpha({\bf K})$ is the so-called  dephasing rate. In this problem, one has that  
\begin{equation}\label{al1}
\Gamma = \frac{\alpha (K)}  {1-\cos(K a \cos(\pi/6))} .
\end{equation}
where the dephasing rate is extracted from the experiment. This rate is not only a function of $K$ but also implicitly of surface temperature and friction coefficient. The total jumping rate, $\Gamma$, is usually plotted versus the inverse of the temperature in an Arrhenius-like plot.\cite{jardine2010determination,Torres2025} 

After Eq.(\ref{ISF2}), the $N^{th}$-derivative of the characteristic function or momentum generating function with respect to $K$ at $K=0$ gives the $N^{th}$-moment of the probability distribution \cite{Gardiner,Haag,Weiss}
\begin{equation}\label{MGF}
\langle X^N(t)\rangle =
\left( - i \frac{\partial}{\partial K } \right)^N I(K,t) \big|_{K=0} ,
\end{equation}
and, similarly, the $N^{th}$-cumulant generating function is given by 
\begin{equation}\label{CGF}
\langle X^N(t)\rangle_c =
\left( - i \frac{\partial}{\partial K } \right)^N \ln I(K,t) \big|_{K=0} .
\end{equation}
This transport process is referred to as diffusive when all moments and cumulants grow linearly with time at long times; for example, in the incoherent surface tunneling.  

\begin{figure}
	\includegraphics[width=0.5\linewidth]{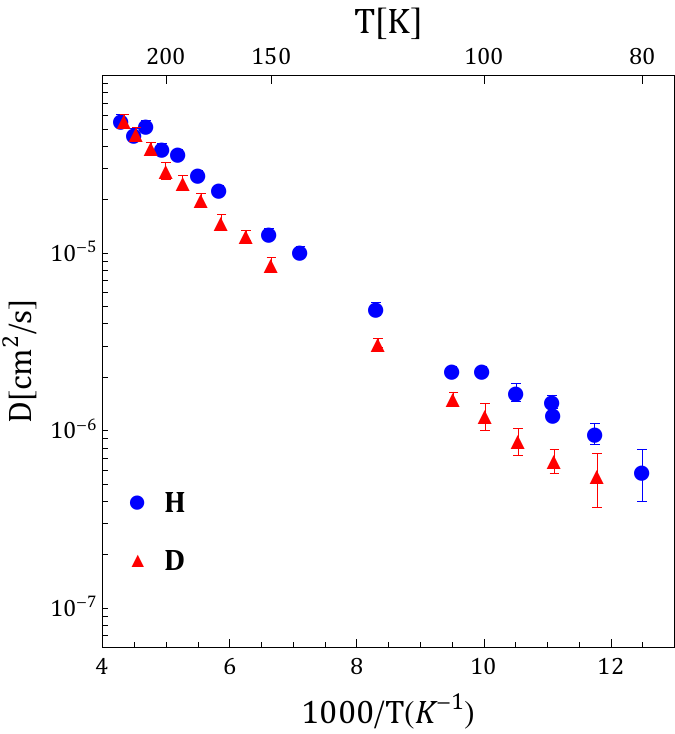}
	\caption{Diffusion coefficient issued from Eq. (\ref{D1}) versus the inverse of the surface temperature for the incoherent tunneling of H and D on a Pt(111) surface. Blue points are for H and, red points, for D. The experimental total hopping/tunneling rates are measured for a momentum transfer of $K= 0.86\, \textup{~\AA}^{-1}$ along the $[11\bar2]$ direction and a coverage of 0.1 ML.}
	\label{F1}
\end{figure}

It is quite straightforward to calculate the first two moments and cumulants for the problem at hands. Thus, one has that 
\begin{eqnarray} 
\langle X(t) \rangle&=&\langle X(t) \rangle_c = 0, \nonumber \\
\langle X^2(t)\rangle &=& \langle X^2(t) \rangle_c = D t, 
\end{eqnarray}
where the diffusion coefficient $D$ is expressed as 
\begin{equation}\label{D1}
D= \Gamma a^2 cos^2 \beta     ,
\end{equation}
which is different from the one used previously.\cite{jardine2010determination}
When the first moment is equal to zero, the three first moments and cumulants are equal. They start differing from the fourth order.

In Fig. \ref{F1},  the diffusion coefficient is plotted as a function of the inverse temperature by considering the experimental total jump rate \cite{jardine2010determination} divided by $2/3$ since only two equivalent direction are active instead of 3 for the direction of observation. The diffusion coefficients are three times greater than those previously reported in the experimental work.

If hops are not restricted to only nearest neighbors, for a simple Bravais lattice and one of the symmetry directions, the corresponding master equation is expressed as
\begin{equation}\label{Pdot2}
	\dot P_n (t) = \sum_{l=1}^{\infty} \, \big[  \Gamma^+_{l} P_{n-l}(t) +  \Gamma^-_{l} P_{n+l} (t) - (   \Gamma^+_{l} + \Gamma^-_{l} )  P_{n}(t) \big]  .
\end{equation}
The ISF can then be written as a product of backward and forward diffusion
\begin{equation}
I(K,t) = \prod_{n=1}^{\infty} I_n^+(K,t) \, I_n^-(K,t)  ,
\end{equation}
with 
\begin{equation}
I_n^{\pm}(K,t) = e^{t \, (e^{\pm i n \, K \, a \, cos \beta }-1) \, \Gamma_n^{\pm}} ,
\end{equation}
where again $a$ and $\beta$ are the unit cell length along the symmetry direction considered and $\beta$ the angle formed by this direction and $\textbf{K}$, respectively. An alternative expression for the ISF can also be written as (from the CE model)
\begin{equation}\label{eq:I}
I({\bf K}, t) = e^{-  2 \sum_{n > 0} \Gamma_n \, [ 1 - cos (n \, K \, a \ cos \beta)]  \, t} ,
\end{equation}
where $\sum_n \Gamma_n = \Gamma$. From this expression, the cumulants of the probability distribution function, $\langle X^N(t) \rangle_c$ are straightforward calculated. Thus, we have that
\begin{eqnarray} 
\langle X(t) \rangle_c&=&\langle X(t) \rangle = 0, \nonumber \\
\langle X^2(t)\rangle_c &=& \langle X^2(t) \rangle = D t, 
\end{eqnarray}
where the diffusion coefficient $D$ is expressed as 
\begin{eqnarray}
D&=& \Gamma a^2 cos^2 \beta   \sum_{n = 1} \bar P_n n^2  ,\nonumber \\
&=& \Gamma  \frac{b^2 + 2 b +2}{b^3} e^{-b} a^2 cos^2 \beta
\end{eqnarray}
where $\bar P_n = \Gamma_n / \Gamma$. Usually, the jump probabilities decreases exponentially with $n$, $\bar P_n= \exp (-b\,n)$, and the summation over $n$ could be replaced by an integral from 1 to infinity (resulting expression written in the second line)

Along this work, we have showed that the experimental (observable) ISF can also be seen as a characteristic function of probability theory. This fact has important consequences since through the  moment and cumulant generating functions, one can access in a simple analytical way to a complete information of the probability distribution function of the adsorbate position in time. In particular, the diffusion coefficient is easily extracted from the second order moment and cumulant since both quantities coincide when the first moment and first cumulant are zero.

%

%

\section*{Acknowledgments}
E.E.T.-M. and S.M.-A. acknowledge support of a grant from the Ministry of Science, Innovation and Universities with Ref. PID2023-149406NB-I00.



\bibliography{rsc} 

\bibliographystyle{rsc.bst}

\end{document}